\documentclass[aps,pra,amsmath,amssymb,floatfix,twocolumn,amsmath,superscriptaddress,twocolumn,nofootinbib,tighten,letterpaper]{revtex4-1}
\usepackage{multirow}
\usepackage{bbold}
\usepackage{subfigure}
\usepackage{color}
\usepackage{mathrsfs}
\usepackage{hyperref}
\usepackage[normalem]{ulem}
\usepackage{bm}

\usepackage{amsfonts, relsize, color}
\usepackage{graphics}
\usepackage{graphicx}
\usepackage{subfigure}
\usepackage{hyperref}
\usepackage{color}
\usepackage{comment}

\renewcommand\vec[1]{\ensuremath\boldsymbol{#1}} 

\begin{document}
\title{Higher-order topological phases on fractal lattices}

\author{Sourav Manna}
\affiliation{Department of Condensed Matter Physics, Weizmann Institute of Science, Rehovot 7610001, Israel}
\affiliation{Max-Planck-Institut f\"{u}r Physik komplexer Systeme, N\"{o}thnitzer Stra{\ss}e 38, 01187 Dresden, Germany}

\author{Snehasish Nandy}
\affiliation{Department of Physics, University of Virginia, Charlottesville, Virginia, 22904, USA}

\author{Bitan Roy}
\affiliation{Department of Physics, Lehigh University, Bethlehem, Pennsylvania, 18015, USA}

\date{\today}
\begin{abstract}
Electronic materials harbor a plethora of exotic quantum phases, ranging from unconventional superconductors to non-Fermi liquids, and, more recently, topological phases of matter. While these quantum phases in integer dimensions are well characterized by now, their presence in fractional dimensions remains vastly unexplored. Here, we theoretically show that a special class of crystalline phases, namely, higher-order topological phases that via an extended bulk-boundary correspondence feature robust gapless modes on lower dimensional boundaries, such as corners and hinges, can be found on a representative family of fractional materials: \emph{quantum fractals}. To anchor this general proposal, we demonstrate realizations of second-order topological insulators and superconductors, supporting charged and neutral Majorana corner modes, on planar Sierpi\'{n}ski carpet and triangle fractals, respectively. These predictions can be experimentally tested on designer electronic fractal materials, as well as on various highly tunable metamaterial platforms, such as photonic and acoustic lattices.    
\end{abstract}

\maketitle

\emph{Introduction}.~Crystals are ubiquitous in nature, manifesting discrete reflection, rotational, and translational symmetries. On the other hand, quasicrystals and fractals are paradigmatic examples of noncrystalline materials. While quasicrystals are projections of higher-dimensional crystals on lower-dimensional branes, realized by completely tilling the physical space in an aperiodic fashion, thereby exhibiting local discrete, often crystal forbidden, rotational symmetries~\cite{janot:QCbook, edited:QCbook, Senechal:QCbook}, fractals by contrast display a fourth type of symmetry, \emph{self-similarity}, resulting in pattern repetition over many scales~\cite{fractal:book}. Fractals appear at macroscale (coastline and trees), as well as at microscales, with the recently engineered electronic Sierpi\'{n}ski triangle in designer materials opening a paradigm of \emph{quantum fractals}~\cite{cmsmith2019:frac}. Despite being embedded in integer $d$-dimensional space, fractals are characterized by irrational Hausdorff or fractal dimension $d_{\rm frac}<d$. Therefore, when combined with the geometry and topology of the electronic wavefunction, quantum fractals give rise to a rich, still vastly unexplored, landscape of topology in fractional dimensions~\cite{wu2015:frac, neupert2018:frac, cmsmith2019:frac, spaiprem2019:frac, katsnelson2020:frac, souravmanna2020:frac, larsfirtz2020:frac, segev2020:frac, souravmanna2021:frac, mannaroy2022:NHfrac}.

\begin{figure}[t!]
\includegraphics[width=0.98\linewidth]{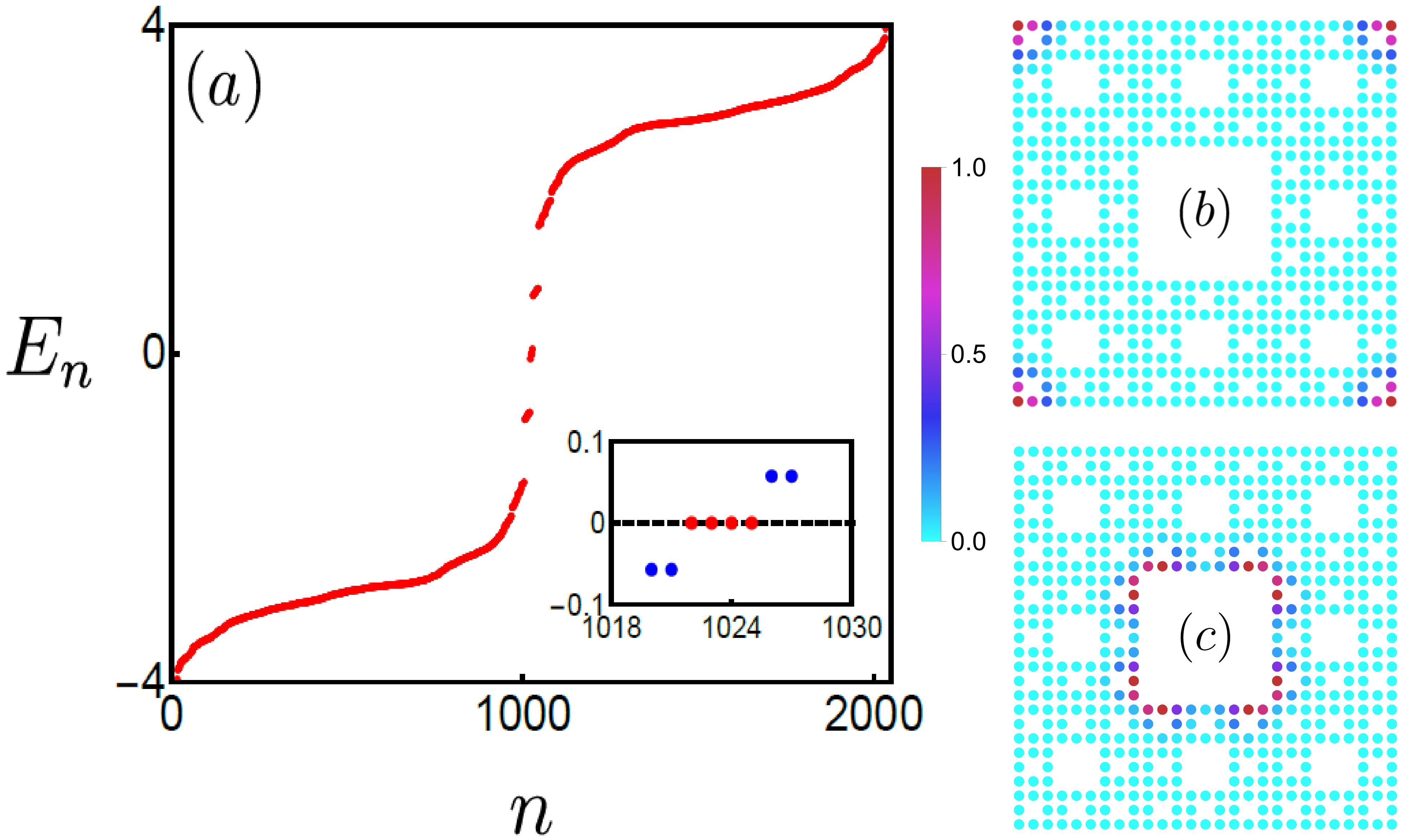}
\caption{HOT insulator on a Sierpi\'{n}ski carpet fractal. (a) Energy spectra of the Hamiltonian $H$ [Eq.~(\ref{eq:HOTIrealspace})] on a Sierpi\'{n}ski carpet fractal of generation $f=3$ (containing 512 sites) for $t=t_0=1$, $m_0=0$ and $g=\sqrt{2}$. (b) The local density of states (LDOS) of the four near zero energy modes, shown in red in the inset of (a), confirms their sharp corner localization. For the definition of the generation number, see Sec.~S1 and Fig.~S1 of the Supplemental Material (SM)~\cite{supplementary}. (c) Localization of four states with closest to zero, but finite energy [blue dots in the inset of (a)] near the innermost corners. See Fig.~S4 of the SM. We normalize LDOS by its maximum value.    
}~\label{fig:carpetspectraLDOS}
\end{figure}

\begin{figure}[t!]
\includegraphics[width=0.98\linewidth]{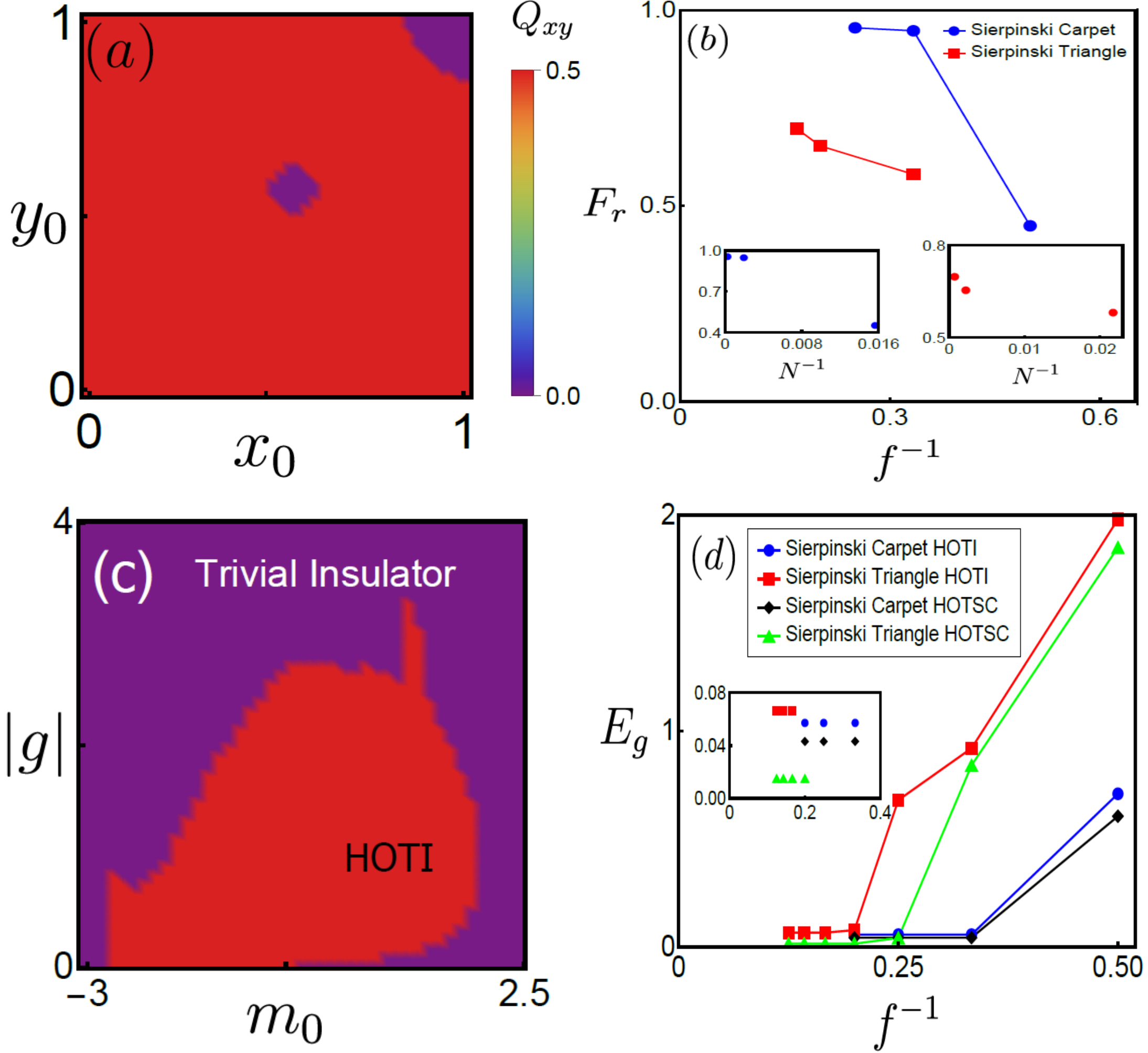}
\caption{(a) Origin $(x_0, y_0)$ dependence of the quadrupole moment $Q_{xy}$ (modulo 1) of a HOT insulator, supporting corner modes (Fig.~\ref{fig:carpetspectraLDOS}) on a Sierpi\'{n}ski carpet fractal of generation $f=3$ containing $N=512$ sites~\cite{supplementary} with open boundary conditions for $t=t_0=1$, $m_0=0$ and $g=\sqrt{2}$. Here $x_0$ and $y_0$ are measured in units of $L$, the linear dimension of the system in each direction. Except for a very few origin choices we indeed find $Q_{xy}=0.5$. (b) Scaling of the fraction of the area $F_r$ in the $(x_0,y_0)$ plane, displaying $Q_{xy}=0.5$, with the inverse of the generation number $f$ and site number $N$ (inset) in the Sierpi\'{n}ski carpet (blue dots) and glued Sierpi\'{n}ski triangle (red squares) fractals for the same parameter values as in (a). In the thermodynamic limit ($f$ or $N \to \infty$) as $F_r \to 1$, $Q_{xy}$ becomes origin independent. (c) Global phase diagram in the $(m_0,|g|)$ plane showing HOT (trivial) insulator with $Q_{xy}=0.5 \; (0.0)$ on Sierpi\'{n}ski carpet fractal for $t=t_0=1$. (d) Scaling of the spectral gap ($E_g$) between the zero energy corner modes and the closest to zero energy modes that are not outer corner localized for HOT insulators (HOTIs) and HOT superconductors (HOTSCs) in two fractal lattices, ensuring that $E_g$ remains finite (inset) in the thermodynamic limit. Here $E_g$ is computed by finding energies of a few states near zero energy using the Lanczos algorithm (not an exact diagonalization).           
}~\label{fig:Qxycarpet}
\end{figure}

\begin{figure}[t!]
\includegraphics[width=0.98\linewidth]{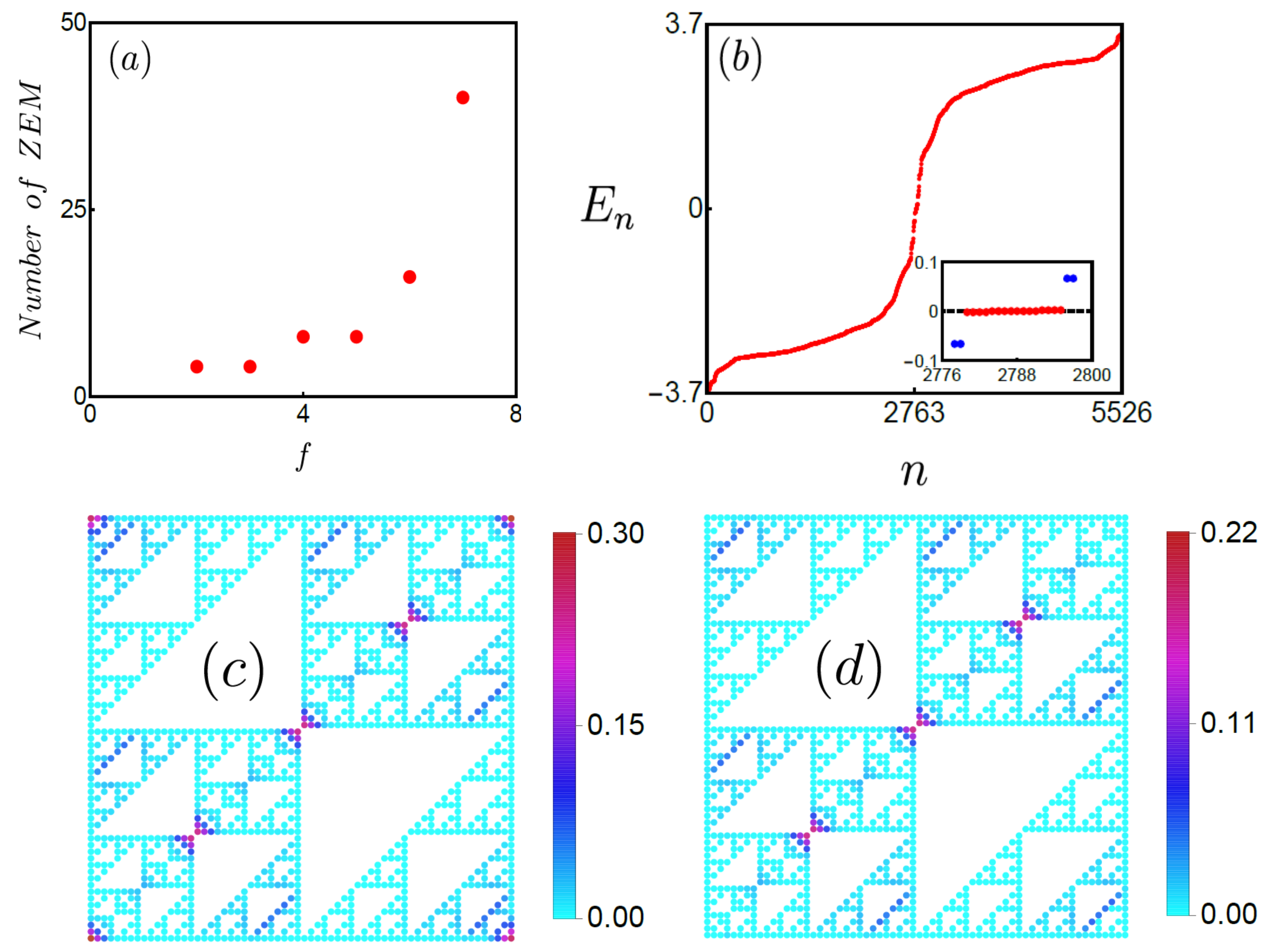}
\caption{HOT insulator on a glued Sierpi\'{n}ski triangle fractal. (a) Number of zero energy modes (ZEM) with the generation number (insensitive to boundary condition). (b) Energy spectra in generation $f=6$ (containing 1394 sites) for $t=t_0=1$, $m_0=0$ and $g=\sqrt{2}$ with open boundary conditions. (c) and (d) Spatial distribution of LDOS for the 16 near-zero-energy modes in a system with open and periodic boundary conditions, respectively.    
}~\label{fig:trianglespectraLDOS}
\end{figure}

Here, we explore this territory by focusing on a newly emerged family of crystalline phases, namely higher-order topological (HOT) phases, and show realizations of both HOT insulators and HOT superconductors on Sierpi\'{n}ski carpet and \emph{glued} Sierpi\'{n}ski triangle fractals (Figs.~\ref{fig:carpetspectraLDOS}-\ref{fig:trianglespectraLDOS}). In general, HOT phases via an extended bulk-boundary correspondence host robust topological modes on lower-dimensional boundaries, such as corners and hinges, characterized by respective codimensions $d_c=d$ and $d-1$~\cite{benalcazar:2017, benalcazar-prb:2017, song:2017, trifunovic:2017, calugaru:2019, fulga:2019, nag-juricic-roy:2019, szabo:2020, wieder-vernevig:2020, xuqiasicrystal:2020, cooper:2020, nag-juricic-roy:2021, trauzettel:2021, gong:2021, bhatbera:2021, gangchen1, gangchen2, bansil:2021, czchen:2021, wang-lin-hughes-HOTSC, wu-yan-huang-HOTSC, liu-he-nori-HOTSC, zhu-HOTSC, pan-yang-chen-xu-liu-liu-HOTSC, ghorashi-HOTSC, broyrantiunitary, trauzettel-HOTSC, dassarma-HOTSC, srao:HOTSC, broysoloHOTSC2020, kheirkhah2020, sigrist2020, tiwari2020, ghoshnagsaha2021, shen2021, bernevig2021, tommyli2021, amundsen-juricic:2021, roy-juricic-octupole:2021}. As such, a HOT phase of order $n$ can be constructed from its conventional first-order counterpart by systematically introducing $n$ number of suitable discrete symmetry breaking Wilson-Dirac masses that partially gap out the edge or surface states, for example, with $d_c=1$, leaving the modes residing on boundaries with $d_c=n$ gapless~\cite{calugaru:2019, nag-juricic-roy:2021}. We show that this principle is operative on fractal lattices as well. In particular, when the global shape of these two fractals is tailored in such a way that four corners reside along the inversion axes of the second-order Wilson-Dirac mass, both HOT insulators and HOT superconductors support robust topological corner modes (Figs.~\ref{fig:carpetspectraLDOS} and \ref{fig:trianglespectraLDOS}). Moreover, the HOT insulators possess quantized quadrupole moment $Q_{xy}=0.5$, which becomes origin independent in the thermodynamic limit, indicating their intrinsic nature (Fig.~\ref{fig:Qxycarpet}). By contrast, $Q_{xy}$ in HOT superconductors exhibit a significant origin dependence, and are thus possibly extrinsic in nature.

The HOT phases on fractals are unique in the sense that they harbor \emph{inner} corner modes, besides the conventional outer corner modes which can also be observed in crystals. However, due to distinct internal geometries such inner corner modes are at finite but close to zero energy (still separated from the rest of the states) in the Sierpi\'{n}ski carpet fractal (Fig.~\ref{fig:carpetspectraLDOS}), while they are pinned at zero energy on the glued Sierpi\'{n}ski triangle fractal (Fig.~\ref{fig:trianglespectraLDOS}).

\emph{Model}.~To outline the general protocol of engineering HOT phases, here we consider its paradigmatic example on a square lattice, captured by the Hamiltonian operator $\hat{h}=\hat{h}_1 + \hat{h}_2$, where
\begin{eqnarray}~\label{eq:hamilHOTImomentum}
\hat{h}_1 &=& t \left[ \sin (k_x a) \sigma_3 \tau_1 + \sin (k_y a) \sigma_0 \tau_2 \right] + M(\vec{k}) \sigma_0 \tau_3, \nonumber  \\
\hat{h}_2 &=& g \left[ \cos(k_x a)-\cos(k_y a) \right] \sigma_1 \tau_1. 
\end{eqnarray}          
The uniform first-order Wilson-Dirac mass 
\begin{equation}
M(\vec{k})= m_0+ 2t_0 - t_0 \left[ \cos(k_x a) + \cos(k_y a) \right]
\end{equation}
preserves all discrete symmetries. Two sets of the Pauli matrices $\{ \sigma_\mu \}$ and $\{ \tau_\mu \}$ operate on the spin and orbital indices, respectively, with $\mu=0,\ldots, 3$. Hereinafter we set the lattice spacing $a=1$. Only in the parameter regime $-2<m_0/t_0<4$, $\hat{h}_1$ features two counter-propagating one dimensional edge modes with $d_c=1$ for opposite spin projections, thereby yielding a first-order quantum spin Hall insulator. Otherwise, the system is a trivial or normal insulator, devoid of any topological edge states~\cite{BHZmodel}.

The second-order Wilson-Dirac mass $\hat{h}_2$ anticommutes with $\hat{h}_1$. It thus acts as a mass to one-dimensional counter propagating edge modes of $\hat{h}_1$ by causing hybridization between them. Naturally, $\hat{h}_2$ gaps out the edge modes, however, only partially as it assumes the profile of a domain-wall mass flipping sign four times under $2 \pi$ rotation and vanishing along the diagonal $\langle 11 \rangle$ directions. Thus $\hat{h}_2$ breaks four-fold rotational ($C_4$) symmetry. As a result, when the corners of a square lattice reside along its diagonals, four corner modes with $d_c=2$ get pinned therein, following the spirit of the generalized Jackiw-Rebbi mechanism~\cite{jackiwrebbi}. We then realize a second-order topological insulator. These modes appear at zero energy due to both unitary and antiunitary particle-hole symmetries of $\hat{h}$, generated by $C=\sigma_2 \tau_1$ and $\Theta=\sigma_3 \tau_1 {\mathcal K}$, respectively, where ${\mathcal K}$ is the complex conjugation, as $\{ \hat{h},C\}=\{ \hat{h},\Theta\}=0$~\cite{broyrantiunitary}. The model also breaks the time reversal symmetry, generated by ${\mathcal T}=\sigma_2 \tau_0 {\mathcal K}$, and parity, generated by ${\mathcal P}=\sigma_0 \tau_3$ under which $\vec{k} \to -\vec{k}$, thus preserving composite $C_4 {\mathcal T}$, $C_4 {\mathcal P}$ and ${\mathcal P} {\mathcal T}$ symmetries.

\emph{Fractal HOT insulators}.~This mechanism is not restricted to the square lattice. If we maintain the symmetry of the model and cleave the system such that four corners are placed along the inversion axes of the HOT Wilson-Dirac mass, it can support corner localized zero-energy modes. To extend the jurisdiction of this model beyond the realm of topological crystals, we consider a real space version of $\hat{h}$, given by $H=H_1+H_2$, with          
\begin{eqnarray}~\label{eq:HOTIrealspace}
		H_{1} &=& \sum_{j \neq k} \frac{G(r_{jk})}{2} c^\dagger_j \bigg[  -i t ( \sigma_3 \tau_1 \cos\phi_{jk} + \sigma_0 \tau_2 \sin\phi_{jk})
\nonumber \\		
		&-& t_0 \sigma_0\tau_3  \bigg]c_k + \sum_{j} c_j^\dagger (m_0 + 2 t_0) \sigma_0 \tau_3 c_j, \nonumber \\
		H_{2} &=& g\sum_{j \neq k} \frac{G(r_{jk})}{2} c^\dagger_j ( \cos 2\phi_{jk}) \sigma_1 \tau_1 c_k,
\end{eqnarray}
and $c_j = [c_{j \uparrow \alpha}, c_{j \uparrow \beta}, c_{j \downarrow \alpha}, c_{j \downarrow \beta} ]^\top$. Here $c_{j \sigma \tau}$ is the electron annihilation operator at site $j$, with spin projection $\sigma=\uparrow, \downarrow$ and on orbital $\tau=\alpha, \beta$. The azimuthal angle between the $j$th and $k$th lattice sites, located at $\vec{r}_j$ and $\vec{r}_k$, respectively, is $\phi_{jk}$, measured with respect to the horizontal direction. For the derivation of Eq.~(\ref{eq:HOTIrealspace}) from Eq.~(\ref{eq:hamilHOTImomentum}) consult Sec.~S2 of the Supplemental Material (SM)~\cite{supplementary}. In order to ensure that the sites in any noncrystalline lattice remain well connected we replace the nearest-neighbor hopping probabilities by a long range one, described by the rotationally invariant function 
\begin{equation}~\label{eq:decayprofile}
	G(r_{jk}) = \exp{\left( 1 - \frac{|\vec{r}_j -\vec{r}_k|}{r_0} \right)}.
\end{equation}
Here $r_0$ is the decay length, typically set to be the nearest-neighbor distance. In principle, this generalized model for a HOT insulator can be implemented on any noncrystalline systems, such as fractals, amorphous materials~\cite{agarwala:octupolar} and quasicrystals~\cite{fulga:2019, xuqiasicrystal:2020, cooper:2020}, as well as on a regular square lattice. Here we focus on the fractal system and scrutinize the possibility of realizing HOT insulators with corner modes on quantum fractals. It should be noted that irrespective of the geometry and internal structure of the system (such as the connectivity among the sites), the above model always enjoys both unitary and antiunitary particle-hole symmetry, now generated by $C_{\rm lat}=\sigma_2 \tau_1 {\mathrm I}_{\ell \times \ell}$ and $\Theta_{\rm lat}=\sigma_3 \tau_1 {\mathrm I}_{\ell \times \ell} \; {\mathcal K}$, respectively, where ${\mathrm I}_{\ell \times \ell}$ is an $\ell$-dimensional identity matrix and $\ell$ is the number of sites in the system.

Results obtained on a Sierpi\'{n}ski carpet fractal with $d_{\rm frac}=\ln(8)/\ln(3) \approx 1.89$ are shown in Fig.~\ref{fig:carpetspectraLDOS}, depicting four near-zero-energy (due to finite system size) modes, which are well separated from the rest of the spectra. For explicit computation of $d_{\rm frac}$ see Sec.~S1 of the SM~\cite{supplementary}. The spatial distribution of the corresponding local density of states (LDOS) shows that these modes are highly localized at four \emph{outer} corners, while the inner corners are devoid of any such mode, in contrast to Ref.~\cite{spaiprem2019:frac}. This observation strongly suggests a possible realization of an electronic HOT insulator on a Sierpi\'{n}ski carpet fractal. Near zero energy there exist four states [blue dots in the inset of Fig.~\ref{fig:carpetspectraLDOS}(a)] that are localized \emph{near} the innermost corners of Sierpi\'{n}ski carpet fractal. See Fig.~\ref{fig:carpetspectraLDOS}(c) and Fig.~S4 of the SM~\cite{supplementary}. Notice that outer corners of the Sierpi\'{n}ski carpet are characterized by the coordination number 2. However, in the interior of the Sierpi\'{n}ski carpet there exists no corner with coordination number 2. Consequently, the blue colored modes from the inset of Fig.~\ref{fig:carpetspectraLDOS}(a) never become zero energy states and their local density of states spreads slightly away from the inner corners.

To anchor this claim, we compute the quadrupole moment $Q_{xy}$ for the fractal HOT insulators~\cite{hughes:octupolar, cho:octupolar, agarwala:octupolar}. To proceed, we first evaluate  
\begin{equation}
n={\rm Re} \left[ -\frac{i}{2 \pi} {\rm Tr} \left( \ln \left\{ U^\dagger  \exp \left[ 2 \pi i \sum_{\bf r} \hat{q}_{xy} ({\bf r}) \right]  U \right\} \right) \right],
\end{equation}
where $\hat{q}_{xy} ({\bf r})= x y \hat{n}({\bf r})/L^2$, $\hat{n}({\bf r})$ is the number operator at ${\bf r}=(x,y)$ of an open boundary system of linear dimension $L$ in each direction, and $U$ is constructed by columnwise arranging the eigenvectors for the negative energy filled states. The quadrupole moment is then defined as $Q_{xy}=n-n_{\rm al}$ (modulo 1), where $n_{\rm al}=(1/2) \; \sum_{\bf r} x y /L^2$ represents $n$ in the atomic limit and at half filling. As each single-particle state is occupied by one particle, the computation of $Q_{xy}$ rests on the fermionic nature of quasiparticles which has no classical analog. Identification of HOT insulators from quantized $Q_{xy}=0.5$ thus justifies the name ``quantum fractal". The results are displayed in Fig.~\ref{fig:Qxycarpet}(a). We compute $Q_{xy}$ for all origin choices. When the HOT insulator supports corner modes, for most of the origin choices $Q_{xy}$ is quantized to $0.5$ within the numerical accuracy. However, in any finite system there always exist a few origin choices for which $Q_{xy}=0$, despite the presence of the corner modes. Such an origin dependence can be quantified by $F_r$, measuring the fraction of all origin choices for which corner modes corroborate quantized $Q_{xy}=0.5$. As the generation number $f$ or number of lattice sites $N$ is increased~\cite{supplementary}, $F_r \to 1$ in the thermodynamic limit, corresponding to $f \to \infty$ or $N \to \infty$ [Fig.~\ref{fig:Qxycarpet}(b)]. The quadrupolar operator $\hat{q}_{xy}(\vec{r})$ is gauge invariant, and the variation of the charge centers ($\vec{r}$) or the origin is tantamount to a gauge transformation~\cite{cho:octupolar}, in turn allowing us to scrutinize the gauge invariance of $Q_{xy}$ when computed in a quantum many-body ground state. Hence, the origin independence of $Q_{xy}$ in the thermodynamic limit ensures it gauge independence, and it stands as a bonafide order parameter for HOT insulators on quantum fractals. Thus, the HOT insulator on a Sierpi\'{n}ski carpet fractal is \emph{intrinsic} in nature.

The ultimate origin independence of $Q_{xy}$ allows us to construct a global phase diagram in the $(m_0,|g|)$ plane [Fig.~\ref{fig:Qxycarpet}(c)]. It supports two topologically distinct phases: (a) a fractal HOT insulator with $Q_{xy}=0.5$ and (b) a trivial insulator with $Q_{xy}=0$. Small and moderate (large) values of $|m_0|$ and $|g|$ are conducive to a HOT (trivial) insulator. Only the entire fractal HOT insulator phase supports four zero energy corner modes. The stability of the fractal HOT insulator can be established from the scaling of the gap between corner modes with the closest finite energy states (not corner localized), shown in red and blue, respectively in the inset of Fig.~\ref{fig:carpetspectraLDOS}(a), with the generation and site numbers. This gap remains finite as we approach the thermodynamic limit [Fig.~\ref{fig:Qxycarpet}(d)], in turn ensuring that corner modes are separated by a \emph{finite} gap, thereby yielding stability to the fractal HOT insulator.

Next we investigate the possibility of realizing HOT insulators on a \emph{glued} Sierpi\'{n}ski triangle fractal. In order to obtain four outer corners along the inversion axes of the second-order Wilson-Dirac mass, we glue two Sierpi\'{n}ski triangle fractals, each being a right angled triangle, slightly different from its known geometry~\cite{fractal:book}. Consequently, the corresponding fractal dimension is $d_{\rm frac}=\ln(6)/\ln(\sqrt{8}) \approx 1.72$ (see Sec.~S1 of the SM~\cite{supplementary}). Numerical diagonalizations reveal that the number of zero energy modes can depend on the generation number [Fig.~\ref{fig:trianglespectraLDOS}(a)]. In the sixth generation there are altogether 16 such modes [Fig.~\ref{fig:trianglespectraLDOS}(b)], well separated from the other nearby states [Fig.~\ref{fig:Qxycarpet}(d)]. As HOT insulators are crystalline topological phases, the number of zero energy modes and their spatial distributions depend on structural details of the system. See, for example Fig.~4 of Ref.~\cite{roy-juricic:dislocation}. On the glued Sierpi\'{n}ski triangle, the number of zero energy modes increases with generation number $f$, as the number of inner naked corners increases with it. However, it always describes the same topological phase, namely, the HOT insulator, characterized by $Q_{xy}=0.5$.

The LDOS of zero-energy modes predominantly occupies four outer corners in a system with open boundaries [Fig.~\ref{fig:trianglespectraLDOS}(c)], qualitatively similar to the situation in a Sierpi\'{n}ski carpet fractal. However, in contrast, the LDOS of all zero energy modes also displays \emph{subdominant} localization at the inner shared \emph{naked} corners, which are devoid of other neighboring sites. Therefore, the manifold of the zero energy modes does not fragment between the outer and inner naked corners. See Fig.~S5 of the SM~\cite{supplementary}. Consequently, in a periodic system, the number of zero energy modes remains unchanged and the corresponding LDOS appears only at the inner corners [Fig.~\ref{fig:trianglespectraLDOS}(d)]. Additionally, the LDOS \emph{weakly} spreads over the inner edges making a $\pi/4$ angle with the horizon, since the Wilson-Dirac mass vanishes in that direction [Figs.~\ref{fig:trianglespectraLDOS}(c), and \ref{fig:trianglespectraLDOS}(d)].

The HOT insulators with outer and naked inner corner modes on glued Sierpi\'{n}ski triangle fractals possess quantized $Q_{xy}=0.5$, which \emph{slowly} becomes origin independent as we approach the thermodynamic limit [Fig.~\ref{fig:Qxycarpet}(b)]. The slowness of $F_r \to 1$ possibly stems from the inner edges at the $\pi/4$ angle, which always absorb a tiny fraction of the LDOS associated with the zero energy modes. The global phase diagram of this system in the $(m_0,|g|)$ plane is qualitatively similar to the one in Fig.~\ref{fig:Qxycarpet}(c). See Fig.~S2 of the SM~\cite{supplementary}.

\emph{Fractal HOT superconductors}.~Continuing the journey through the territory of HOT phases on quantum fractals, next we search for HOT superconductors on Sierpi\'{n}ski carpet and glued Sierpi\'{n}ski triangle fractals. In principle, with suitable choices of Hermitian matrices and the corresponding spinor, which includes both electron and hole like components (Nambu doubling), $\hat{h}$ can also describe a second-order topological superconductor [Eq.~(\ref{eq:hamilHOTImomentum})]. Namely, the quantity appearing with $t$ describes an odd parity $p$-wave pairing, the term proportional to $g$ represents an even parity $d_{x^2-y^2}$ pairing, and $M(\vec{k})$ gives rise to a Fermi surface when $-2<m_0/t_0<4$ on a square lattice with only nearest-neighbor hopping amplitude. The resulting mixed parity, time-reversal symmetry breaking $p+id$ pairing is a prominent candidate for a HOT superconductor that supports four corner localized Majorana zero modes~\cite{wang-lin-hughes-HOTSC, broysoloHOTSC2020}. Naively, it is, therefore, tempting to conclude that quantum fractals harbor HOT superconductors based on the results shown in Figs.~\ref{fig:carpetspectraLDOS}-\ref{fig:trianglespectraLDOS}; this conclusion, however, encounters a few fundamental as well as practical shortcomings.

Primarily, the Hamiltonian $\hat{h}$ does not reveal any microscopic origin of the $p+id$ pairing nor does it unveil any potential material platform where such pairing can be realized. Even more importantly, when we extend $\hat{h}$ to a real space hopping Hamiltonian [Eq.~(\ref{eq:HOTIrealspace})], the pairing terms (proportional to $t$ and $g$) become infinitely long-ranged connecting all the sites with decaying amplitude of the Cooper pairs [Eq.~(\ref{eq:decayprofile})], which is unphysical. Finally, the notion of a Fermi surface in the absence of an underlying translational symmetry, as in fractals, becomes moot. To circumvent these limitations we search for a suitable material platform where \emph{on site} or \emph{local} pairings can give rise to HOT superconductors, which do not strictly rely on a sharp Fermi surface. A class of systems that satisfies all these realistic requisite features is the second-order Dirac insulator, whose normal state is described by the Hamiltonian $\hat{h}$ [Eq.~(\ref{eq:hamilHOTImomentum})]. To accommodate superconducting orders in this system, we Nambu double the spinor. The Hamiltonian then reads as $\hat{h}_{\rm Nam}=\eta_3 \hat{h}_1 + \eta_0 \hat{h}_2$. The newly introduced Pauli matrices $\{ \eta_\mu \}$ with $\mu=0, \ldots, 3$ operate on the Nambu or particle-hole index. Here, we focus only on the local or on site pairings which are oblivious to the underlying lattice structure, and thus possess natural immunity against the lack of crystalline order. Due to the Pauli exclusion principle, the number of such pairings is restricted to be \emph{six}, which is exactly the number of purely imaginary four-dimensional Hermitian matrices. See Sec.~S4 of the SM~\cite{supplementary} for details.

The local second-order topological superconductor can be unambiguously identified from its requisite symmetries. For example, it must anticommute with the Dirac kinetic energy, captured by the terms proportional to $t$ in $\hat{h}_{\rm Nam}$, such that the pairing represents a topological Nambu-Dirac mass. In addition, it must commute with the first-order Wilson-Dirac mass, so that the boundary modes of this pairing are not uniformly gapped. Finally, it must anticommute with the second-order Wilson-Dirac mass such that the Majorana edge modes are gapped, but only partially, producing localized zero energy Majorana modes at four corners, when they reside along the $\langle 11 \rangle$ directions. These constraints select a unique candidate for the second-order topological superconductor, for which the effective single particle Bogoliubov de-Gennes Hamiltonian reads 
\begin{equation}
\hat{h}_{\rm pair}= \Delta \left( \eta_1 \cos \phi + \eta_2 \sin \phi \right) \sigma_1 \tau_2. 
\end{equation}   
Here $\Delta$ is the pairing amplitude and $\phi$ is the U(1) superconducting phase. The Nambu Hamiltonian $\hat{h}^{\rm total}_{\rm Nam}=\hat{h}_{\rm Nam}+\hat{h}_{\rm pair}$ can be implemented on any fractal lattice following Eq.~(\ref{eq:HOTIrealspace}). Without loss of generality, we set $\phi=0$.

The resulting energy spectra and LDOS corresponding to the near zero energy modes are qualitatively similar to the ones shown in Figs.~\ref{fig:carpetspectraLDOS} and~\ref{fig:trianglespectraLDOS} on the Sierpi\'{n}ski carpet and glued Sierpi\'{n}ski triangle fractals, respectively. See Fig.~S3 of the SM~\cite{supplementary}. These observations confirm the realization of HOT superconductors on quantum fractals. Furthermore, to attribute the resulting corner modes solely to the paired state, we choose the normal state to be topologically trivial. However, the quadrupole moment associated with a second-order topological superconductor is found to be $Q_{xy}=0.5$ for a very few origin choices and there is no clear indication of $F_r \to 1$ in the thermodynamic limit, due to strong interband scattering. Therefore, in all likelihood the fractal HOT superconductors, in contrast to their insulating counterparts, are \emph{extrinsic} in nature. Still the spectral gap between (near) zero energy corner modes and other closest to zero energy (not corner localized) states approaches a finite value in the thermodynamic limit [Fig.~\ref{fig:Qxycarpet}(d)]. So, extrinsic fractal HOT superconductors and their hallmark corner modes are stable. These outcomes remain qualitatively unaltered even when the normal state is a fractal HOT insulator.

\emph{Summary and discussions}.~Here, we construct a concrete path to theoretically harness HOT phases on a family of fractional materials, \emph{quantum fractals}, and demonstrate their realizations on Sierpi\'{n}ski carpet and glued Sierpi\'{n}ski triangle fractals. While the HOT insulators are intrinsic in nature, their superconducting cousins are possibly extrinsic. Nonetheless, the HOT paired state in a second-order Dirac insulator is energetically most favored among all symmetry allowed local pairings over a wide parameter range~\cite{supplementary}. This procedure can be generalized to identify HOT phases on fractals with different geometries, as well as on higher-dimensional fractals~\cite{fractal:book, souravmanna2020:frac, marcopolini}. Furthermore, by stacking planar HOT fractals in the out of plane direction one can construct HOT semimetals in a hybrid dimension. These exciting possibilities, inhabiting the landscape of topological quantum fractals, will be systematically explored in the future following our general principle of construction.

Electronic fractal materials, such as the ones recently engineered in designer electronic~\cite{cmsmith2019:frac} and molecular~\cite{wu2015:frac} compounds, constitute the ideal platform where our proposed fractal HOT insulators and superconductors can be realized in experiments. In these quantum fractals, while the insulating HOT phases can be unveiled by designing appropriate hopping elements, their pairing counterparts should become energetically favored upon chemical doping. Our predicted fractal HOT insulators can also be tailored on various classical metamaterials, such as photonic~\cite{rechtsman:2018} and phononic or acoustic~\cite{huber:2018, bahl:2018} lattices, with longer range coupling between the photonic waveguides and microwave resonators, respectively. Topolectric circuits constitute yet another promising platform where our predictions can be tested~\cite{thomalecircuit:2018, dong-juricic-roy:2021}, especially given that quasicrystalline quadrupole insulators have already been realized therein~\cite{xucircuit:2021}, as well as HOT insulators with long range hopping~\cite{topolectric:longrange}. For practical purposes, it should be noted that it is not necessary for the hopping amplitudes to be sufficiently long ranged [Eq.~(\ref{eq:decayprofile})]. As long as all the sites on fractal lattices stay connected, all our findings remain qualitatively unchanged. Although topological boundary modes in classical metamaterials can be detected from the spatial distribution of on-resonance impedance (topolectric circuits) or two-point pump probe spectroscopy (photonic lattices) or absorption spectra (phononic lattices), many-body quantum topological invariants, such as the quadrupole moment $Q_{xy}$, cannot be measured in these systems.

\emph{Note added}.~Recently, we became aware of two experimental works~\cite{sierHOTmeta1Exp, sierHOTmeta2Exp}, where our predictions of HOT insulators in Sierpi\'{n}ski carpet fractals have been observed.

\emph{Acknowledgments}.~S.M. thanks the Weizmann Institute of Science, Israel, Deans Fellowship through the Feinberg Graduate School for financial support.~S.N. acknowledges NSF Grant No.~DMR-1853048. B.R. was supported by a startup grant from Lehigh University.


\end{document}